# Multi-terminal electrical transport measurements of molybdenum disulphide using van der Waals heterostructure device platform

Xu Cui[1‡], Gwan-Hyoung Lee[2‡*], Young Duck Kim[1], Ghidewon Arefe[1], Pinshane Y. Huang[3], Chul-Ho Lee[4], Daniel A. Chenet[1], Xian Zhang[1], Lei Wang[1], Fan Ye[5], Filippo Pizzocchero[6], Bjarke S. Jessen[6], Kenji Watanabe[7], Takashi Taniguchi[7], David A. Muller[3,8], Tony Low[9], Philip Kim[10], and James Hone[1*]

[1]Department of Mechanical Engineering, Columbia University, New York, NY 10027, USA

[2]Department of Materials Science and Engineering, Yonsei University, Seoul 120-749, Republic of Korea

[3]School of Applied and Engineering Physics, Cornell University, Ithaca, NY 14853, USA

[4]KU-KIST Graduate School of Converging Science and Technology, Korea University, Seoul 136-701, Republic of Korea

[5]Department of Material Science and Engineering, Columbia University, New York, NY 10027, USA

[6]Center for Nanostructured Graphene (CNG), Technical University of Denmark, DK-2800 Kongens Lyngby, Denmark

[7]National Institute for Materials Science, 1-1 Namiki, Tsukuba 305-0044, Japan

[8]Kavli Institute at Cornell for Nanoscale Science, Ithaca, NY 14853, USA

[9]Department of Electrical & Computer Engineering, University of Minnesota, Minneapolis, MN 55455, USA

[10]Department of Physics, Harvard University, Cambridge, MA 02138, USA

‡These authors contributed equally.

Corresponding authors: gwanlee@yonsei.ac.kr and jh2228@columbia.edu

**Atomically thin two-dimensional (2D) semiconductors such as molybdenum disulphide (MoS$_2$) hold great promise in electrical, optical, and mechanical devices[1–4] and display novel physical phenomena such as coupled spin-valley physics and the valley Hall effect[5–9]. However, the electron mobility of mono- and few-layer MoS$_2$ has so far been substantially below theoretically predicted limits[10–12], particularly at low temperature (*T*), which has hampered efforts to observe its intrinsic quantum transport behaviors. Potential sources of disorder and scattering include both defects such as sulfur vacancies in the MoS$_2$ itself, and extrinsic sources such as charged impurities and remote optical phonons from oxide dielectrics[10,11,13,14]. To reduce extrinsic scattering and approach the intrinsic limit, we developed a van der Waals (vdW) heterostructure device platform where MoS$_2$ layers are fully encapsulated within hexagonal boron nitride (hBN), and electrically contacted in a multi-terminal geometry using gate-tunable graphene electrodes. Multi-terminal magneto-transport measurements show dramatic improvements in performance, including a record-high Hall mobility reaching 34,000 cm$^2$/Vs for 6-layer MoS$_2$ at low *T*. Comparison to theory shows a decrease of 1-2 orders of magnitude in the density of charged impurities, indicating that performance at low *T* in previous studies was limited by extrinsic factors rather than defects in the MoS$_2$[15–17]. We also observed Shubnikov-de Haas (SdH) oscillations for the first time in high-mobility monolayer and few-layer MoS$_2$. This novel device platform therefore opens up a new way toward measurements of intrinsic properties and the study of quantum transport phenomena in 2D semiconducting materials.**

Following the many advances in basic science and applications of graphene, other 2D materials, especially transition metal dichalcogenides (TMDCs), have attracted significant interest for their fascinating electrical, optical, and mechanical properties[3,4,16,18–22]. Among the TMDCs, semiconducting $MoS_2$ has been the mostly widely studied: it shows a thickness-dependent electronic band structure[18,20], reasonably high carrier mobility[3,4,15,16,21,22], and novel phenomena such as coupled spin-valley physics and the valley Hall effect[5–8,23], leading to various applications, such as transistors[3,22,24], memories[25], logic circuits[26,27], light-emitters[28], and photo-detectors[29] with flexibility and transparency[4,30]. However, as for any 2D material, the electrical and optical properties of $MoS_2$ are strongly affected by impurities and its dielectric environment [3,4,15,31], hindering the study of intrinsic physics and limiting the design of 2D-material-based devices. In particular, the theoretical upper bound of the electron mobility of single-layer (1L) $MoS_2$ is predicted to be from several tens to a few thousands at room $T$ and exceed $10^5$ cm$^2$/Vs at low $T$ depending on the dielectric environment, impurity density and charge carrier density[10–12]. In contrast, experimentally measured 1L $MoS_2$ devices on $SiO_2$ substrates have exhibited room-$T$ two-terminal field-effect mobility that ranges from 0.1 - 55 cm$^2$/Vs[3,32,33]. This value increases to 15 - 60 cm$^2$/Vs with encapsulation by high-$k$ materials[3,15], owing to more effective screening of charged impurities[11]. Due to the presence of large contact resistance from the metal-$MoS_2$ Schottky barrier, however, these two-terminal measurements underestimate the true channel mobility[22,34,35]. Multi-terminal Hall mobility measurements[15,16] still show mobility substantially below theoretical limits, particularly at low $T$ with best reported values of 174 cm$^2$/Vs at 4 K for 1L[15] and 250 cm$^2$/Vs and 375 cm$^2$/Vs at 5 K for 1L and 2L [16]. Typically, these thin samples exhibit a crossover to non-metallic behavior at carrier densities below ~$10^{13}$ cm$^{-2}$ [15,16,36], or at smaller carrier densities by engineering of local defect states and improving interfacial quality[13]. The scattering and disorder that leads to this non-metallic behavior can arise from multiple origins such as lattice defects, charged impurities in the substrate and surface adsorbates, and it has been difficult to identify their separate contributions[3,10,13,15,16,36–38].

We have previously demonstrated that encapsulation of graphene within hBN reduces scattering from substrate phonons and charged impurities, resulting in band transport behavior that is near the ideal acoustic phonon limit at room $T$, and ballistic over more than 15 $\mu$m at low $T$[39]. These results were realized with a novel technique to create one-dimensional edge contacts to graphene exposed by plasma-etching a hBN/graphene/hBN stack. Such an approach has not yet proved effective with $MoS_2$. However, recent reports that graphene can create a high quality electrical contact to $MoS_2$[27,40] motivate a hybrid scheme, in which the channel $MoS_2$ and multiple graphene 'leads' are encapsulated in hBN, and the stack is etched to form graphene-metal edge contacts. This new scheme is distinct from previous approaches, in that the entire $MoS_2$ channel is fully encapsulated and protected by hBN, and that we achieve multi-terminal graphene contacts without any contamination from device fabrication process. In samples fabricated with this approach, we observed ultra-high low-$T$ Hall mobility up to ~ 34,000 $cm^2$/Vs and report the observation of Shubnikov-de Haas (SdH) oscillations, indicating quantization of electron dynamics under high magnetic fields. These results demonstrate a new pathway toward harnessing the intrinsic properties of 2D materials and open up new vistas in exploration of low-$T$ quantum physics of TMDCs.

Fig. 1a and 1b show a schematic diagram and optical micrograph of a Hall bar device structure. We employed a 'PDMS (Polydimethylsiloxane) transfer' technique[4] to place few-layer graphene flakes around the perimeter of an $MoS_2$ flake, encapsulate them between thicker hBN layers, and place the entire stack on a $Si/SiO_2$ wafer. (Fig. S1a) The stack was then shaped into a Hall bar geometry such that hBN-encapsulated $MoS_2$ forms the channel. In the contact regions, graphene overlaps the $MoS_2$ and extends to the edge, where it is in turn contacted by metal electrodes[39]. Details of the fabrication process are described in the Methods section and Supplementary Information S1. High-resolution scanning transmission electron microscopy (STEM) (Fig. 1c; see Fig. S1b for a larger clean interface area of > 3 $\mu$m) confirms that the stacking method can produce ultraclean interfaces free of residue that can be seen when an organic polymer film is used for stacking[41]. We note that while Ohmic contacts have also been achieved in metal-$MoS_2$ contacts by deposition of small work-function metals, vacuum annealing, and electrostatic gating[4, 17, 18], top-deposited metal electrodes are not compatible with hBN-encapsulation.

For this study, a series of samples with thickness from 1 - 6 layers (1L - 6L) was fabricated and measured. The number of layers was identified by Raman and photoluminescence (PL). (See Supplementary Information S2) All samples were obtained by exfoliation except for the 1L sample, for which we used chemical vapor deposition (CVD) grown monolayer $MoS_2$ because of the limited size of exfoliated monolayers. The CVD-grown $MoS_2$ single crystal has been shown to exhibit high quality from structural, electrical and optical measurements[42], although the process of transferring it from the growth substrate may introduce more contamination than for exfoliated flakes.

For each sample, we performed temperature-dependent two-probe measurements to examine the quality of the graphene contacts. Fig. 2a shows output curves ($I_{ds} - V_{ds}$) of a 4L $MoS_2$ device at $V_{bg}$ = 80 V. The response is linear at room $T$ and remains linear to low $T$, indicating an Ohmic contact. Similar behavior is seen for $V_{bg}$ > 20 V, whereas gapped behavior corresponding to non-Ohmic contact is seen for $V_{bg}$ < 20 V. This is consistent with previous studies which show a gate-tunable contact barrier between graphene and $MoS_2$[27,40]. In addition, it establishes the gate voltage range over which multi-terminal measurements can be reliably performed. Fig. 2b shows the measured four-terminal resistivity $\rho$ (in log scale) of the same sample from $V_{bg}$ = 20 V to 80 V (corresponding to carrier densities of ~ $4.8\times10^{12}$ cm$^{-2}$ to ~ $6.9\times10^{12}$ cm$^{-2}$, respectively), and from room $T$ to 12 K. $\rho$ decreases with increasing $V_{bg}$, as expected for *n*-type conduction. With decreasing temperature, $\rho$ drops dramatically over the entire accessible range of $V_{bg}$, reaching 130 $\Omega$ at 12 K. All of the samples studied exhibited similar behavior: *n*-type conduction and metallic temperature-dependence in the gate voltage accessible to four-terminal measurements.

By comparing the two- and four-terminal results, the contact resistance can be determined (see Supplementary Information S3). The results for the 4L MoS$_2$ device, as shown in Fig. 2c, directly demonstrate that the contact resistance can be tuned by back gate voltage. In fact, a small contact resistance of ~ 2 k$\Omega\cdot\mu$m can be reliably achieved at large gate voltage at room *T*. This likely reflects primarily the graphene-MoS$_2$ junction resistance, since both the graphene resistance and the graphene-metal contact resistance should be substantially less[39]. Below $V_{bg}$ = 20 V, the contact resistance increases upon cooling, indicating activated transport across a contact barrier. However, above $V_{bg}$ = 20 V, the contact resistance decreases upon cooling, reaching a low-*T* value of ~ 1 k$\Omega\cdot\mu$m above $V_{bg}$ = 50 V. This metallic behavior directly demonstrates that low-resistance contacts, with no thermal activation, can be achieved at sufficiently high gate voltage. Similar behavior was observed in all samples (Fig. S3), with contact resistance at high $V_{bg}$ ranging from ~ 2 - 20 k$\Omega\cdot\mu$m at room *T* and ~ 0.7 - 10 k$\Omega\cdot\mu$m at low *T*. These values are comparable to room-*T* values reported previously for graphene[46] and metal[43–45] contacts, but larger than the best contacts achieved by MoS$_2$ phase engineering (0.2 - 0.3 k$\Omega\cdot\mu$m)[35].

To examine the quality of the hBN-encapsulated devices and determine the scattering mechanisms limiting the carrier mobility of MoS$_2$, the Hall mobility $\mu_{Hall}(T)$ was derived from $\rho(T)$ and the carrier density $n(V_{bg})$ (obtained by Hall effect measurements, see Supplementary Information S4). Fig. 3a shows $\mu_{Hall}$ for the 1L - 6L samples as a function of temperature, at carrier densities varying from $4.0 \times 10^{12}$ cm$^{-2}$ to $1.2 \times 10^{13}$ cm$^{-2}$ (see Fig.3b and Table 1). For all of the samples, mobility increases with decreasing temperature and saturates at a constant value at low $T$. The low-$T$ mobility in our devices is much higher than previously reported values, and there is no sign of metal-insulator transition as observed at similar carrier densities around $10^{13}$ cm$^{-2}$ in SiO$_2$-supported MoS$_2$[15,16,36–38]. This strongly suggests that extrinsic scattering and disorder (either from SiO$_2$ or from processing with polymer resists) has been the primary source of non-metallic behavior in MoS$_2$ measured to date.

The measured mobility curves can be well fitted to a simple functional form: $\frac{1}{\mu(T)} = \frac{1}{\mu_{imp}} + \frac{1}{\mu_{ph}(T)}$, where $\mu_{imp}$ is the contribution from impurity scattering, and $\mu_{ph}$ is the temperature-dependent contribution due to phonon scattering. From the fitting to the experimental mobility curves in Fig. 3a, we found that $\mu_{ph}(T)$ is well described by a power law $\mu_{ph} \sim T^{-\gamma}$. (Fig. S7) This behavior is consistent with mobility limited by MoS$_2$ optical phonons, as theoretically predicted to have an exponent of ~ 1.69 in monolayer[10] and ~ 2.5 for bulk MoS$_2$[47] at T > 100 K. Although this power law behavior has been observed in experiments by other groups[15,16,36], a stronger temperature dependence was observed in our devices, with the exponent $\gamma$ ranging from 1.9 - 2.5 (inset table of Fig. 3a), as opposed to 0.55 - 1.7 reported previously[15,16]. We also note that the room-$T$ mobility, which is dominated by phonon scattering in all of the samples, is seen to vary from 40 - 120 cm$^2$/Vs, with no discernible trend with thickness, nor a significant variation with carrier density. The physical origin for this variation remains unclear, but disorder or material quality can be ruled out due to the high low-$T$ mobility in all of these samples.

The data shown in Fig. 3a yield values of $\mu_{\mathrm{imp}}$ ranging from 1,020 cm$^2$/Vs in the CVD-grown monolayer to 34,000 cm$^2$/Vs for 6L, up to two orders of magnitude higher than previously reported values[15–17] (Table 1). Because free charges in the material can screen impurities in a thickness-dependent manner, it is important to measure the density-dependent mobility to rigorously compare sample quality, and to validate theoretical models. As shown in Fig. 3b, each sample shows an increase in $\mu_{\mathrm{imp}}$ with increasing carrier density, as expected due to screening. For more quantitative understanding of the effect of interfacial Coulomb impurities on carrier mobility of MoS$_2$, we employed a model based on a perturbative approach by Stern[48], from which we obtained the screened Coulomb potential used in the mobility calculation. This model has also been commonly used in the context of semiconductor devices (see Supplementary Information S6). Within the model, increasing carrier density enhances screening of the interfacial Coulomb potential, which leads to improved carrier mobility, and, increasing the thickness of MoS$_2$ redistributes the charge centroid further from the interface, resulting in enhancement of mobility. The dashed lines in Fig. 3b indicate the simulated curves by using this model, in which the density of interfacial impurities ($D_{it}$), assumed to be distributed equally at the two MoS$_2$/hBN interfaces, is used as a fitting parameter. The key trends of increasing mobility with carrier density and thickness are in large part corroborated by our experiments. The model also allows for more direct comparison of performance across samples – for example, the 1L mobility here follows a curve shifted approximately one decade above previous measurements [16]. The estimated values of $D_{it}$ in our devices are order of $10^9 - 10^{10}$ cm$^{-2}$ (Table 1), 1-2 orders of magnitude smaller than in previously reported devices[15–17]. The good fit to the model across multiple samples with only one fitting parameter also suggests that, even in these extremely clean devices, interfacial impurities remain the dominant scattering mechanism. Therefore further improvements in mobility can be achieved by obtaining even cleaner interfaces without requiring improvements to the quality of the $MoS_2$ source material.

Fig. 4 shows the longitudinal ($R_{xx}$) and Hall resistance ($R_{xy}$) of the 6L (Fig. 4a) and monolayer (Fig. 4b) samples as a function of applied magnetic field ($B$, up to 9 T for 6L, 31 T for 1L). The high carrier mobility in these samples allows us to observe pronounced SdH oscillations for the first time, providing additional strong evidence of high quality and homogeneity in our devices. The quantum mobility ($\mu_Q$), which is limited by both small and large angle scatterings that destroy quantized cyclotron orbit motions, can be estimated from the magnetic field corresponding to the first discernible oscillation, following the relation $\mu_Q \sim 1/B_q$[49]. The quantum mobilities of 1L and 6L $MoS_2$ are ~ 1,400 $cm^2$/Vs and ~ 10,000 $cm^2$/Vs, respectively, in line with the measured Hall mobilities. We note that the positive magneto-resistance background signal in $R_{xx}$ and the complicated oscillation behavior of 6L $MoS_2$ are likely due to multi-subband occupation[50]. In contrast, the 1L $MoS_2$ shows more regular oscillations, indicating conduction in a single subband. Encouragingly, the high-field Hall resistance (green curve, $R_{xy}$) begins to reveal plateau-like structures at high magnetic field coinciding with $R_{xx}$ minima. These emerging features were similarly observed in early studies of graphene samples with moderate mobility[51], giving hope that fully developed quantum Hall effect can be observed with further improvements in sample quality.

In conclusion, we demonstrate a vdW heterostructure device platform in which an ultra-thin $MoS_2$ layer is encapsulated by hBN and contacted by graphene. The vdW heterostructure provides a standard device platform that enables us to measure intrinsic electrical transport of 2D materials and achieve high mobility 2D devices for studying the unique transport properties and novel quantum physics. By forming robust and tunable electrical contacts and dramatically reducing Coulomb scattering impurities, ultra-high mobility of $MoS_2$ can be achieved up to approaching the theoretical limit at low *T*. As with graphene, we believe the ability to study the quantum transport and new physics in TMDCs will spark greater interest in the community, allowing for more in-depth studies in the future.

**Methods**

**Sample fabrication.** The $hBN/MoS_2/graphene/hBN$ stacks were fabricated using the 'PDMS transfer'[4] technique on 285 nm $SiO_2/Si$ substrates. The transfer techniques are described in detail in the Supplementary Information S1. The stacks were then shaped to the desired Hall bar structure through electron-beam patterning and reactive ion etching (RIE) with a mixture of $CHF_3$ and $O_2$. Finally, metal leads were patterned by e-beam lithography and subsequent deposition of metals (Cr 1nm/Pd 20nm/Au 50nm). The metal leads make edge-contact to graphene electrodes as reported previously[39].

**TEM sample preparation.** For high-resolution imaging, we fabricated a cross-sectional TEM lift-out sample from the finished encapsulated devices, using a FEI Strata 400 dual-beam Focused Ion Beam. STEM imaging was conducted in a FEI Tecnai F-20 STEM operated at 200kV, with a 9.6 mrad convergence semiangle and high-angle annular dark field detector. False coloring was added by hand.

**Electrical measurements and magneto-transport measurements.** Two-terminal transport characteristics were measured by applying DC bias (Keithley 2400) to the source and gate electrodes and measuring the drain current using a current amplifier (DL 1211). For four-terminal measurements, a standard lock-in amplifier (SR830) measured voltage drop across the channel with constant current bias. Magneto-transport measurements were done in a Physical Property Measurement System (PPMS) (Fig. 4a) and a $He_3$ cryostat at the National High Magnetic Field Laboratory (NHMFL) (Fig. 4b).

**Acknowledgements**

This research was supported by the U.S. National Science Foundation (DMR-1122594), and in part by the FAME Center, one of six centers of STARnet, a Semiconductor Research Corporation program sponsored by MARCO and DARPA. G.H.L was supported by Basic Science Research Program (NRF-2014R1A1A1004632) through the National Research Foundation (NRF) funded by the Korean government Ministry of Science, ICT and Future Planning. P.Y.H. acknowledges support from the NSF Graduate Research Fellowship Program under grant DGE-0707428. Additional support was provide through funding and shared facilities from the Cornell Center for Materials Research NSF MRSEC program (DMR-1120296). F.P. and B.S.J. acknowledged the Center for Nanostructured Graphene (CNG) which is funded by the Danish National Research Foundation, Project DNRF58. The high magnetic field measurements were performed at the NHMFL and the authors thank Alexey Suslov, Bobby Joe Pullum, Jonathan Billings, and Tim Murphy for assistance with the experiments at NHMFL.

**Author Contributions**

X.C. and G.H.L. designed the research project and supervised the experiment. X.C., G.H.L., Y.D.K., G.A., C.H.L., F.Y., F.P., B.S.J., and L.W. performed device fabrication and X.C., G.H.L. and Y.D.K. performed device measurements under supervision of P.K. and J.H.. X.C., G.H.L., G.A., X.Z. performed optical spectroscopy and data analysis. D.A.C. grew and prepared the CVD $MoS_2$ sample. T.L. performed the theoretical calculations. K.W. and T.T. prepared hBN samples. P.Y.H. and D.A.M. performed TEM analyses. X.C., G.H.L. and J.H. analyzed the data and wrote the paper.

**Figure Captions**

**Figure 1 | vdW device structure and interface characterization. a,** Schematic of the hBN-encapsulated MoS$_2$ multi-terminal device. Exploded view shows the individual components that constitute the heterostructure stack. The bottom panel shows the zoom-in cross-sectional schematic of metal-graphene-MoS$_2$ contact region. **b,** Optical microscope image of a fabricated device. Graphene contact regions are outlined by dashed lines. **c,** Cross-section STEM image of the fabricated device. The zoom-in false-color image clearly shows the ultra-sharp interfaces between different layers. (graphene: 5L, MoS$_2$: 3L, top-hBN: 8 nm, bottom-hBN: 19 nm)

**Figure 2 | Gate-tunable and temperature-dependent graphene-MoS$_2$ contact. a,** Output curves ($I_{ds}$ - $V_{ds}$) of the hBN-encapsulated 4L MoS$_2$ device with graphene electrodes at varying temperature. The back gate voltage ($V_{bg}$) is kept at 80 V with carrier density of $6.85 \times 10^{12}$ cm$^{-2}$ in MoS$_2$. The linearity of output curves confirms that graphene-MoS$_2$ contact is Ohmic at all temperatures. **b,** Resistivity of 4L MoS$_2$ (log scale) as a function of $V_{bg}$ at varying temperature. The resistivity decreases upon cooling, showing metallic behavior, reaching ~130 Ω at 12 K. **c,** Contact resistance of the same device as a function of $V_{bg}$ at varying temperature. The inset shows the contact resistance as a function of temperature at different $V_{bg}$. At high $V_{bg}$, contact resistance even decreases when decreasing the temperature.

**Figure 3 | Temperature and carrier density dependence of Hall mobility. a,** Hall mobility of hBN-encapsulated MoS$_2$ devices with different number of layers of MoS$_2$ as a function of temperature. To maintain Ohmic contacts, a finite $V_{bg}$ was applied. The measured carrier densities from Hall measurements for each device are listed in Table 1. The solid fitting lines are drawn by the model in the main text. All the fitting parameters are listed in Table 1. For a visual guideline, a dashed line of power law $\mu_{ph}$ ~ T$^{-\gamma}$ is drawn and fitted values of $\gamma$ for each device are listed in the

inset table. **b,** Impurity-limited mobility ($\mu_{imp}$) as a function of carrier density of $MoS_2$. For comparison, the previously reported values from $MoS_2$ on $SiO_2$ substrates (Ref. 16, 17) are plotted. The dashed lines show theoretical calculations of $\mu_{imp}$ as a function of carrier density (*n*) at certain impurity density ($D_{it}$) which extracted and listed in Table 1.

**Figure 4 | Observation of Shubnikov-de Hass oscillations in hBN-encapsulated MoS₂ device.**
**a,** Longitudinal resistance $R_{xx}$ and Hall resistance $R_{xy}$ of hBN-encapsulated 6L $MoS_2$ device as a function of magnetic field (*B*). Hall measurement was conducted at 3 K and at $V_{bg}$ = 80 V, corresponding to carrier density of $5.32 \times 10^{12}$ cm$^{-2}$. **b,** $R_{xx}$ and $R_{xy}$ of hBN-encapsulated CVD 1L $MoS_2$ device as a function of *B* measured at 0.3 K at $V_{bg}$ = 100 V (carrier density of $9.69 \times 10^{12}$ cm$^{-2}$) and oscillation starts around 7.2 T. SdH oscillation is clearly observed in CVD 1L and 6L $MoS_2$ samples and $R_{xy}$ shows plateau-like behavior at $R_{xx}$ minima.

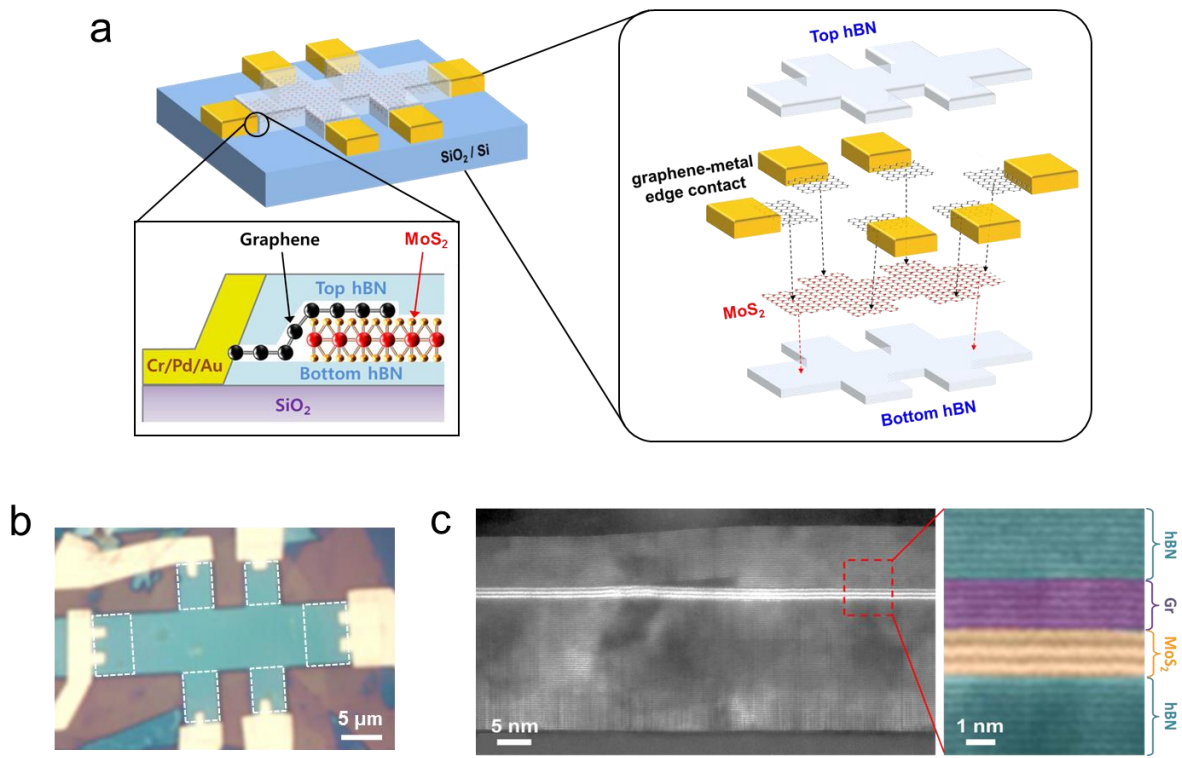

**Fig. 1**

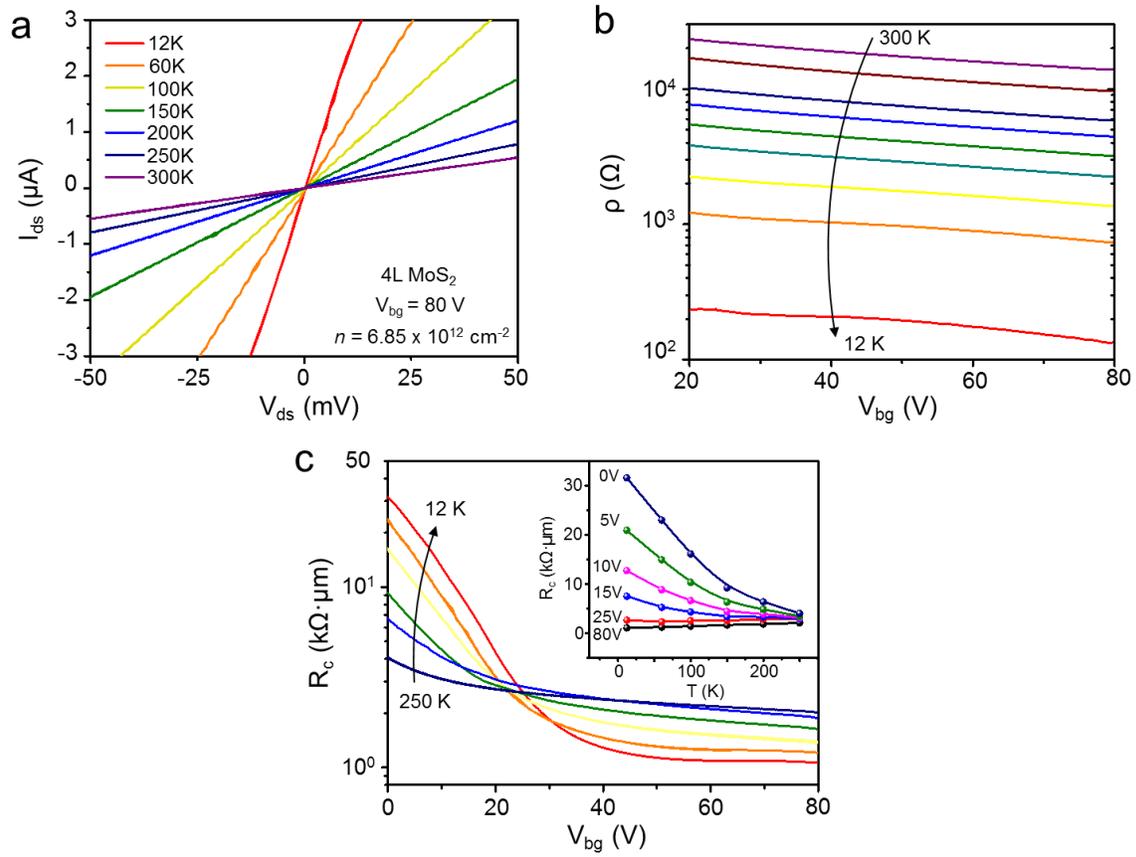

**Fig. 2**

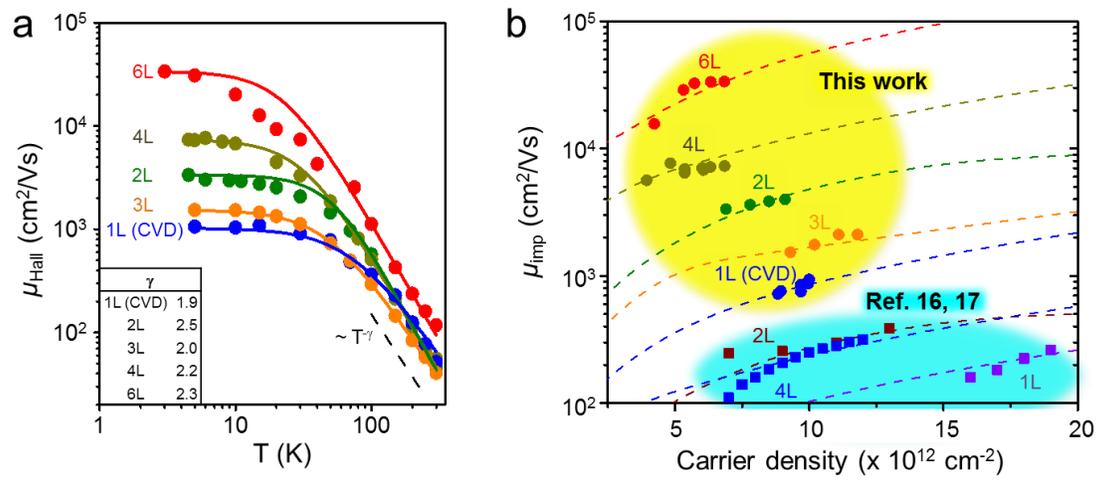

**Fig. 3**

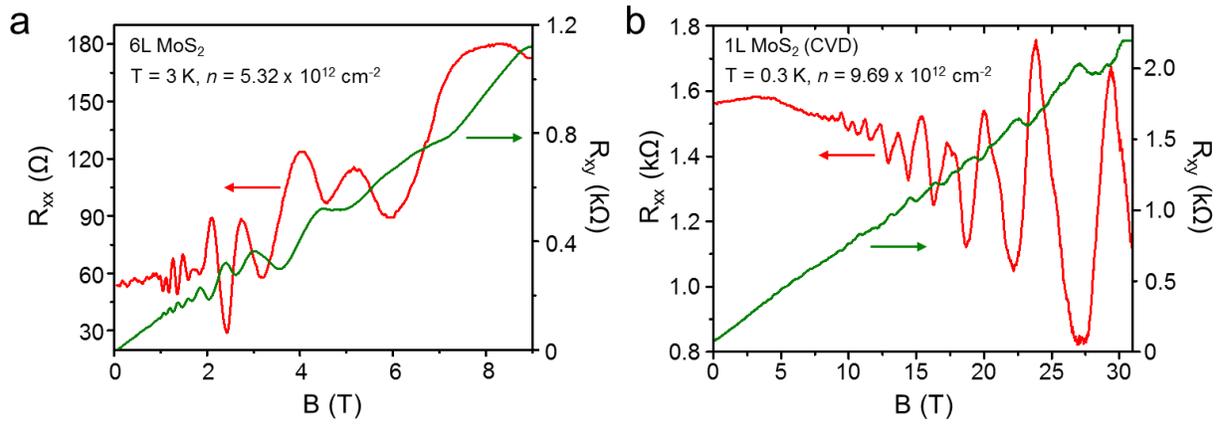

**Fig. 4**

| Layer Number | $D_{it}$ (cm$^{-2}$) | $\mu_{imp}$ (cm$^2$/Vs) | $n$ (cm$^{-2}$) | $\gamma_{ph}$ |
|---|---|---|---|---|
| 1L (CVD) | 6x10$^{10}$ | 1,020 | 1.1x10$^{13}$ | 1.9 |
| 2L | 1.7x10$^{10}$ | 3,350 | 6.9x10$^{12}$ | 2.5 |
| 3L | 3.7x10$^{10}$ | 1,530 | 9.3x10$^{12}$ | 2.0 |
| 4L | 4x10$^9$ | 7,300 | 6.9x10$^{12}$ | 2.2 |
| 6L | 1x10$^9$ | 34,000 | 6.9x10$^{12}$ | 2.3 |
| 1L (Ref.16) | 5x10$^{11}$ | 264 | 1.9x10$^{13}$ | 1.7 |
| 2L (Ref.16) | 3x10$^{11}$ | 394 | 1.3x10$^{13}$ | 1.1 |
| 4L (Ref.17) | 2x10$^{11}$ | 314 | 1.2x10$^{13}$ | 1.5 |

**Table 1**